\documentclass[aps,prd,twocolumn,groupedaddress,showpacs,nofootinbib,amssymb]{revtex4-2}
\usepackage[dvips]{graphicx}
\usepackage{amssymb}
\usepackage{amsmath}
\usepackage{graphicx,,color}
\usepackage{amsfonts}
\usepackage{bm}
\usepackage{cancel}
\usepackage{comment}

\newcommand\be{\begin{equation}}
\newcommand\ee{\end{equation}}

\DeclareUnicodeCharacter{2212}{-}
\allowdisplaybreaks[4]

\begin{document}

\tolerance=5000

\title{Reheating era in Gauss-Bonnet theories of gravity \\
compatible with the GW170817 event}
\author{S.A. Venikoudis\, \thanks{venikoudis@gmail.com}}
\author{F.P. Fronimos\,\thanks{fotisfronimos@gmail.com}}
\affiliation{
 Department of Physics, Aristotle University of
Thessaloniki, Thessaloniki 54124,
Greece
}

\tolerance=5000

\begin{abstract}
In the present article we showcase how the reheating era can be described properly in the context of Einstein-Gauss-Bonnet gravity assuming that the primordial gravitational waves propagate with the velocity of light. The equations of the duration of reheating along with the reheating temperature are derived and as demonstrated, their expressions are quite similar to the case of a canonical scalar field where now the second derivative of the Gauss-Bonnet scalar coupling function appears and effectively alters the numerical value of the scalar potential. The appearance of such term is reminiscing of a $\lambda R$ model of gravity where $\lambda$ is now dynamical. We consider two viable inflationary models of interest, the former involves an error function as scalar Gauss-Bonnet coupling function and the latter a Woods-Saxon scalar potential. It is shown that for both models the aforementioned quantities can be in agreement with theoretical expectations. The only constraint that is needed is the assumption that the second time derivative of the Gauss-Bonnet scalar coupling function is actually lesser than the Planck mass squared, that is $\ddot\xi<\frac{M_{Pl}^2}{8}$ in order to obtain a viable description. We find that a free parameter of the theory and specifically the potential amplitude for the Woods-Saxon model during the inflationary era, dictates the effective equations of state and therefore the reheating epoch can be described either by a type of stiff matter with EoS parameter equal to unity or by an EoS parameter close to that of the radiation. 
\end{abstract}


\maketitle

\section{Introduction}
The cosmological evolution during the primordial Universe constitutes one of the most intriguing challenges in modern Cosmology. According to the paradigm which was firstly introduced in the pioneer paper \cite{Guth:1980zm}, the inflationary era succeeds the quantum epoch and the Universe expands with a quasi-exponential rate. During this era the supercooling process occurs where, due to the abrupt expansion, the temperature of the Universe drops at low values. The reheating era succeeds inflation while the Universe still expands with lower rate and the temperature starts to increase in order to reach the pre-inflationary values. This era provides a mechanism, which is necessary for the cosmological evolution given that the matter must be thermalized so as the radiation dominated era can follow. 
The description of the cosmological dynamics during the reheating era is quite a challenging task for many reasons, with the most important of them being the lack of experimental evidence, meaning that the numerical values of  the reheating temperature and the reheating duration have yet to be determined from observations. 

Until now, many alternative scenarios about the description of the reheating mechanism have been proposed in the literature see, Ref. \cite{Oikonomou:2017bjx,Mijic:1986iv,Cook:2015vqa,Mathew:2020nqo,vandeBruck:2016xvt,Koh:2018qcy,Watanabe:2006ku,Ellis:2021kad,Rinaldi:2015uvu,Amin:2014eta,Choi:2016eif,Fei:2017fub,Martin:2014nya,Gong:2015qha,Cai:2015soa,Ueno:2016dim,Eshaghi:2016kne,DeHaro:2017abf,Artymowski:2017pua,Opferkuch:2019zbd,Dimopoulos:2018wfg,Lankinen:2019ifa,Hasegawa:2019jsa,Sadjadi:2012zp,Braden:2010wd,Nakayama:2008wy,Harigaya:2013vwa,Tashiro:2003qp,Kofman:2005yz,Munoz:2014eqa,Dai:2014jja,Allahverdi:2010xz} and recently the theory where an abnormal reheating epoch can be produced from higher curvature terms on the energy spectrum of primordial gravitational waves, see Ref. \cite{Odintsov:2022sdk}. We approach the reheating era as semi-classical considering the Einstein's gravitational background with the involvement of Gauss-Bonnet string corrective term. The motivation for this reasoning arises from the fact that
the effective Lagrangian of inflation has not been determined yet. Thus, we assume that the imprint of the quantum era remains on the Lagrangian even though the Planck era has been ceased.

The main goal of the work is the construction of a theoretical framework in Einstein-Gauss-Bonnet gravity which describes the reheating era of the primordial Universe, having been preceded by the inflationary era, driven by a canonical scalar field. In the following, the structure of the paper is briefly demonstrated. In Sect.II the reheating temperature and the duration of the reheating era are extracted as functions of the scalar potential. In addition, by imposing the constraint that tensor perturbations propagate through spacetime with the velocity of light, we extract
the form of an auxiliary function which involves the Gauss-Bonnet scalar coupling function and affects the reheating process.
In Sect.III the inflationary dynamics of the GW170817 compatible Einstein-Gauss-Bonnet model are discussed. Finally, in Sect.IV two inflationary models are discussed where the first is described by an error function Gauss-Bonnet scalar coupling function while the latter by the Woods-Saxon scalar potential, with the purpose of extracting information about the reheating era for the same set of parameters. For the sake of generality we consider two possible scenarios for the potential amplitude of the second model, which determines the allowed EoS parameter for the reheating era.

As a last note before we commence the study it should be stated that throughout this paper the cosmological background will be assumed to be flat and homogeneous with the line element being equal to,
\begin{equation}
\centering
\label{metric}
\mathrm{d}s^2=-\mathrm{d}t^2+a^2(t)\delta_{ij}\mathrm{d}x^i\mathrm{d}x^j\, ,
\end{equation}
with $a(t)$ being the scale factor. Furthermore, the reduced Planck mass $M_{Pl}$ is assumed to be equal to unity however, it is kept in the following expressions for the sake of consistency.

\section{Reheating Era In Gauss-Bonnet Gravity}

In order to extract information about the cosmological dynamics during the reheating era of the Universe, let us first specify the gravitational action of the theory. Here, we shall assume that the action contains only a canonical scalar field which is non-minimally coupled to the Gauss-Bonnet topological invariant, that is
\begin{equation}
\centering
\label{action}
\mathcal{S}=\int{\mathrm{d}^4x\sqrt{-g}\left(\frac{R}{2\kappa^2}-\frac{1}{2}g^{\mu\nu}\nabla_\mu\phi\nabla_\nu\phi-V(\phi)-\xi(\phi)\mathcal{G}\right)}\, ,
\end{equation}
where $R$ denotes the Ricci scalar, $\kappa=\frac{1}{M_{Pl}}$ with $M_{Pl}$ being the reduced Planck mass, $\mathcal{G}=R_{\mu\nu\sigma\rho}R^{\mu\nu\sigma\rho}-4R_{\mu\nu}R^{\mu\nu}+R^2$ stands for the Gauss-Bonnet topological invariant with $R_{\mu\nu\sigma\rho}$ and $R_{\mu\nu}$ being the Riemann and Ricci tensor respectively and finally $V(\phi)$ and $\xi(\phi)$ are the scalar potential and the Gauss-Bonnet scalar coupling function of the model. In principle, these two scalar functions represent the two degrees of freedom of the gravitational model however, we shall showcase subsequently that they are interconnected. It should also be stated that due to the fact that the line element coincides with the one from the Friedmann-Robertson-Walker metric, it is reasonable to limit our work on a homogeneous canonical scalar field, meaning that the scalar field is only time dependent, namely $\phi=\phi(t)$.

Before we proceed with the reheating era, it is important to note that the gravitational waves in this primordial era are assumed to propagate with the velocity of light. In general, this is a statement that needs to be highlighted in the context of Gauss-Bonnet theories of gravity, since it is not valid for each Gauss-Bonnet scalar coupling function. If one studies the propagation of tensor perturbations, according to the Ref. \cite{Hwang:2005hb}, it becomes apparent that their velocity is given by the following expression,
\begin{equation}
\centering
\label{ct}
c_\mathcal{T}^2=1-\frac{Q_f}{Q_t}\, ,
\end{equation}
where $Q_f=8(\ddot\xi-H\dot\xi)$ and $Q_t=\frac{1}{\kappa^2}-8\dot\xi H$ are auxiliary functions and the dot as usual implies differentiation with respect to the cosmic time $t$. Hence, it becomes abundantly clear that in its current form, the field propagation velocity of tensor perturbations is time dependent and satisfies the condition $c_T^2\le 1$ in Natural units. Since we require that the primordial gravitons remain massless during the inflationary and the reheating era, we conclude that the tensor perturbations propagate with the velocity of light once the Gauss-Bonnet scalar coupling function satisfies the differential equation $\ddot\xi=H\dot\xi$. This condition serves as a powerful constraint during the inflationary era that manages to transform the continuity equation of the scalar field to a first order differential equation of the scalar field, for details see Ref. \cite{Odintsov:2020sqy} or subsequent sections. As a result, the effective degrees of freedom decrease and now a simple designation of the Gauss-Bonnet scalar coupling function suffices, in order to draw conclusions about the scalar potential and the evolution of the scalar field and vice versa. Hence, the degrees of freedom remain exactly the same as for the case of a canonical scalar field regardless of the appearance of the non-minimal coupling.

Let us now focus on the reheating era and in particular in its duration and the reheating temperature. In order to obtain information about such quantities, the equations of motion of the theory must be extracted from the gravitational action (\ref{action}). By varying the action with respect to the metric tensor and the scalar field, the following set of equations is generated,
\begin{equation}
\centering
\label{motion1}
 \frac{3H^2}{\kappa^2}=\frac{1}{2}\dot\phi^2+V+24\dot\xi H^3\, ,
\end{equation}

\begin{equation}
\centering
\label{motion2}
-\frac{2\dot H}{\kappa^2}=\dot\phi^2-16\dot\xi H\dot H\, ,
\end{equation}

\begin{equation}
\centering
\label{motion3}
\ddot\phi+3H\dot\phi+V'+\xi'\mathcal{G}=0\, ,
\end{equation}
where for simplicity, differentiation with respect to the canonical scalar field is indicated with the prime. These equations are valid throughout the evolution of the Universe however, we shall modify them a bit in order to describe the reheating era. Firstly, let us denote the function $X(\phi)=8\ddot\xi(\phi)$ for later convenience and introduce a transfer term in the continuity equation of the scalar field, given that the radiation dominated era succeeds the reheating era. In this case, the equations of motion should be altered as,

\begin{equation}
\centering
\label{motion4}
H^2=\frac{\kappa^2}{3(1-\kappa^2X)}\left(\frac{1}{2}\dot\phi^2+V\right)\, ,
\end{equation}

\begin{equation}
\centering
\label{motion5}
\dot H=-\frac{\kappa^2\dot\phi^2}{2(1-\kappa^2X)}\, ,
\end{equation}

\begin{equation}
\centering
\label{motion6}
\ddot\phi+(3H+c_0\Gamma)\dot\phi+\xi'\mathcal{G}=0\, ,
\end{equation}
where $\Gamma=\frac{g^4\sigma^2}{8\pi m_\phi}+\frac{h^2m_\phi}{8\pi}$ as usual serves as the decay rate of the scalar field to bosons (first term) and fermions (second term) and the couplings are assumed to be of order $\mathcal{O}(10^{-3})$ based on the Ref.\cite{vandeBruck:2016xvt}. For generality we include the parameter $c_0=0,1$ which acts as a switch for enabling interactions between the scalar field and radiation. Notice that the expression of the Hubble parameter and its derivative are greatly simplified due to the constraint $\ddot\xi=H\dot\xi$ on the velocity of tensor perturbations. In consequence, the slow-roll index $\epsilon_1=-\frac{\dot H}{H^2}$ has the same form as in the case of the canonical scalar field given by the condition $\xi=0$. This means that in the inflationary era, the slow-roll condition $\kappa^2X\ll1$ for the Gauss-Bonnet scalar coupling function is not required in order to simplify the first slow-roll index but, can be used on the Gauss-Bonnet scalar coupling function in order to simplify the expression of the scalar potential (\ref{motion3}). In general one may not consider such condition but make use of the usual slow roll conditions, meaning $\dot H\ll H^2$, $\frac{1}{2}\dot\phi^2 \ll V$ and $\ddot\phi\ll H\dot\phi$. 

In order to extract the expression of the duration of reheating era, we shall make use of the definition of the e-folding number $N=\ln a$ where $a$ denotes the scale factor. In this case, the reheating era commences during the ending stage of inflation and ceases during the initial stage of the radiation dominated era, therefore knowledge of the scale factor at such time instances specifies completely the duration of reheating, that is $N_{re}=\ln\left(\frac{a_{re}}{a_f}\right)$ where $a_{re}$ and $a_f$ is the scale factor during the final stage of reheating and inflation respectively. To facilitate the study, it is convenient to replace scale factors with energy densities since one can operate with energy densities more freely given that the equations of motion are not only known but also simplified due to the aforementioned constraint. Recalling that the energy density for a perfect fluid scales as,

\begin{equation}
\centering
\label{endensity}
\rho=\rho_0a^{-3(1+\omega)}\, ,
\end{equation}
with $\omega$ being the equation of state (EoS) parameter at a given cosmological era, the e-folding number in the reheating era obtains the following form,

\begin{equation}
\centering
\label{efolds1}
N_{re}=\frac{1}{3(1+\omega_{re})}\ln\left(\frac{\rho_f}{\rho_{re}}\right)\, ,
\end{equation}
with the EoS parameter of the reheating era being in the vicinity of $-\frac{1}{3}<\omega_{re}\leq1$, between the final value of the effective EoS in the inflationary era and stiff matter. It is interesting to state that while the equation of state shall be assumed to be constant throughout the reheating era, the value of $\omega_{re}=-\frac{1}{3}$ shall not be considered since it describes neither acceleration nor deceleration given that $\ddot a=0$. The case of stiff matter however shall be studied since it is the maximally allowed value for the EoS that respects causality. Overall, this form is quite convenient since the energy density at such time instances is known. Firstly, since the ending of reheating coincides with the start of the radiation dominated era, it becomes apparent that the relativistic particles, namely photons, are in thermal equilibrium with the following energy density,

\begin{equation}
\centering
\label{photons}
\rho_{re}=\frac{\pi^2}{30}g_*T^4_{re}\, ,
\end{equation}  
where $g_*$ quantifies the relativistic degrees of freedom in the radiation dominated era and is approximately $g_*\sim100$. In order to compute $\rho_f$ from Friedmann's equation, let us first examine how the condition $\epsilon_1(\phi_f)=1$ is translated in terms of the scalar functions. According to equations (\ref{motion4}) and (\ref{motion5}), one can easily observe that by demanding $H^2(\phi_f)=-\dot H(\phi_f)$, the following relation can be extracted,
\begin{equation}
\centering
\label{equivalence}
\dot\phi^2(\phi_f)=V(\phi_f),\, 
\end{equation}
therefore, the energy density when the inflationary era ends has the form,
\begin{equation}
\centering
\label{finalendens}
\rho_f=\frac{3V(\phi_f)}{2(1-\kappa^2X(\phi_f))}.\
\end{equation}
Substituting  Eq. (\ref{photons}) and Eq. (\ref{finalendens}) into Eq. (\ref{efolds1}), one can easily observe that the duration of the reheating era is given by the following expression,
\begin{equation}
\centering
\label{efolds2}
N_{re}=\frac{1}{3(1+\omega_{re})}\ln\left(\frac{45V(\phi_f)}{\pi^2g_*T_{re}^4(1-\kappa^2 X(\phi_f))}\right).\
\end{equation}
It is worth mentioning that the result seems to be quite similar to the one extracted in Ref. \cite{Cook:2015vqa} for the canonical scalar field, the only difference lies on the inclusion of the second time derivative of the Gauss-Bonnet scalar coupling function in the denominator, inside the argument of the logarithm, that is evaluated during the initial stage of the reheating era. Therefore, the correct limit is safely reached for $\xi(\phi)\to0$. In general, the constraint Gauss-Bonnet model seems to be reminiscing of a term $(1-\kappa^2X)R$ in the gravitational action, however now $X$ is dynamical and not constant. In addition, we stress that in Eq. (\ref{efolds2}) the reheating temperature participates. Ideally, we would like to have a separate expression for the e-folding number $N_{re}$ specified by the free parameters of the model and afterwards the above relation can be used in order to find the reheating temperature as a function of the e-folding number. An important issue that should be addressed here is the order of magnitude of the second derivative of the Gauss-Bonnet scalar coupling function. As shown in Eq.(\ref{efolds2}), the difference $1-\kappa^2X(\phi_f)$ participates in the denominator. Since each object is positively defined, the aforementioned difference must also be positive. This suggests that the Gauss-Bonnet scalar coupling function must satisfy the inequality $\ddot\xi<\frac{M_{Pl}^2}{8}$ for the sake of consistency, otherwise the e-folding number is ill defined. It should also be stated that the condition is easily modified in order to account for the possibility of a  modified gravity model of the form $\lambda R$ with $\lambda$ now being a dimensionless parameter. In this case, one finds that the second derivative of the Gauss-Bonnet scalar coupling function should satisfy the condition $\ddot\xi<\frac{\lambda M_{Pl}^2}{8}$.

In order to isolate the e-folding number, let us use the fact that the reheating temperature is connected to the current temperature of the CMB due to entropy conservation as shown below,
\begin{equation}
\centering
\label{retemp}
T_{re}=T_0\frac{a_0}{a_{re}}\left(\frac{43}{11g_*}\right)^{\frac{1}{3}},\, 
\end{equation}
where subscript zero is used in order to label the current values of the temperature and the scale factor. Following the formalism developed in Ref. \cite{Cook:2015vqa}, the final expression of the e-folding number and the reheating temperature are showcased below,
\begin{equation}
\centering
\label{reefolds}
N_{re}=\frac{4}{1-3\omega_{re}}\left[61.6-N_{k}-\ln\left(\frac{1}{H_k}\left(\frac{V(\phi_f)}{1-\kappa^2X(\phi_f)}\right)^{\frac{1}{4}}\right)\right],\,
\end{equation}

\begin{equation}
\centering
\label{Tre}
T_{re}=\left(\frac{45V(\phi_f)}{\pi^2g_*(1-\kappa^2X(\phi_f))}\right)^{\frac{1}{4}}e^{-\frac{3(1+\omega_{re})}{4}N_{re}}\, ,
\end{equation}
where $N_{k}$ indicates the e-folding number that specifies the duration of the inflationary era, estimated to be around $N_{k}=50-60$ and the Hubble rate during the first horizon crossing is given by the expression,
\begin{equation}
\centering
\label{Hk}
H_k=\frac{\pi}{\kappa}\sqrt{8A_k\left|\left(\epsilon_1(\phi_k)-\frac{\kappa^2 Q_e(\phi_k)}{4H(\phi_k)}\right)\frac{c_A^3(\phi_k)}{\kappa^2Q_t(\phi_k)}\right|}\, ,
\end{equation}
as a function of the tensor to scalar ratio (to be presented in the subsequent section). Here, $\phi_k$ is the initial value of the scalar field during the first horizon crossing and $A_k$ denotes the amplitude of curvature perturbations which, according to the Planck 2018 collaboration \cite{Planck:2018jri}, has the numerical value of $1.90461\cdot10^{-9}$. In principle, its exact value does not affect dramatically the overall phenomenology however, its order of magnitude is important and should be kept as $\mathcal{O}(10^{-9})$. We also adopt the notation frequently which is used in literature and use subscript $k$ to describe magnitudes when $k$ modes pass through the horizon for the first time.

At this stage, let us make a brief comment on the e-folding number and the reheating temperature in equations (\ref{reefolds}) and (\ref{Tre}). The extracted results are essentially identical to the case of a canonical scalar field with the only addition being an effective shift in the value of the input of the logarithm in the expression of the e-folding number. Now instead of the scalar potential, the auxiliary parameter $X(\phi)$ also participates in the denominator. In consequence, the fact that the e-folding number is a real value constraints the sign of $X$ effectively. In order to obtain a viable description, one should keep in mind that,
\begin{equation}
\centering
\label{constraint}
\ddot\xi<\frac{M_{Pl}^2}{8}.\
\end{equation}
Note that this inequality, unlike the slow-roll conditions which are implemented in the inflationary era as order of magnitudes, is now a constraint on the sign of $X$. In other words, one could in principle obtain a viable description for $|X|\sim\mathcal{O}(100)$ but only if $X<0$. On the other hand, if $|X|<1$ the both positive and negative values of $X$ are allowed. This condition is important for the computation of the value of $X$ in the final stage of the inflationary era as we shall see subsequently. Similarly, by capitalising on the fact that $\ddot\xi=H\dot\xi$, one can obtain a similar constraint for $H\dot\xi$.

Before we proceed with the numerical results of the duration of the reheating era and the reheating temperature for given values of the reheating EoS parameter for models compatible with the Planck data, it is important to specify the expression of parameter $X(\phi)$. In the previous relations, it was shown that this parameter is equal to $X(\phi)=8\ddot\xi(\phi)$ however, it is time dependent whereas all the rest parameters in Eq.(\ref{reefolds}) and (\ref{Tre}) are scalar field dependent. It would be ideal to find an expression that replaces such time dependence with the scalar field. In our case, such expression is dictated by the propagation velocity of primordial tensor perturbations and reads,
\begin{equation}
\centering
\label{X1}
X=8H\dot\xi\, ,
\end{equation}
for an arbitrary value of the scalar field. This relation seems to carry a time dependence again but in our current formalism, only the numerical value during the ending stage of the inflationary era is desirable, therefore the expression $X(\phi_f)$ as a function of the scalar field needs to be extracted. At such value, it is known that the relation $\dot\phi^2(\phi_f)=V(\phi_f)$ holds true, so the apparent time dependence can be replaced. Another replacement that is also necessary is that $\dot\xi=\xi'\dot\phi$, hence one can extract a simple algebraic relation for $X(\phi_f)$ as shown below,
\begin{equation}
\centering
\label{X2}
\kappa^2X(\phi_f)=\pm\frac{4\sqrt{2}\kappa^3V(\phi_f)\xi'(\phi_f)}{\sqrt{1-\kappa^2X(\phi_f)}}\, ,
\end{equation}
therefore at best, the solution for $X(\phi)$ is real and unique. In order to ensure that the solution is indeed real, the inequality $\kappa^2X(\phi_f)<1$ suffices, as in the case of the e-folding number in Eq. (\ref{reefolds}). It is also worth mentioning that the sign of $\dot\phi_f$, which is essentially specified by the inflationary model, does not affect the overall phenomenology. Furthermore, one may choose to work with the equivalent form for factor $X$, that is $X(\phi)=8\ddot\xi(\phi)$ and expand the second time derivative as $\ddot\xi=\ddot\phi\xi'+\dot\phi^2\xi''$ and essentially replace $\ddot\phi$ from the continuity equation of the scalar field (\ref{motion6}). This approach shall not be considered here since it is more intricate than the one in Eq. (\ref{X1}) however, it should be stated at this point that in case the continuity equation of the scalar field is studied algebraically for $\phi=\phi_f$, the contribution from the Gauss-Bonnet scalar coupling function vanishes in Eq.(\ref{motion6}) since $\mathcal{G}(\phi_f)=0$ due to the fact that $\epsilon_1(\phi_f)=1$.

\section{Inflationary Phenomenology in Einstein-Gauss-Bonnet gravity and compatibility with 
GW170817}
In this section we shall briefly review the inflationary dynamics in the context of Einstein-Gauss-Bonnet gravity, which is compatible with the GW170817 event \cite{LIGOScientific:2017vwq,LIGOScientific:2017ync,LIGOScientific:2017zic,LIGOScientific:2018cki,LIGOScientific:2018hze,Ezquiaga:2017ekz,Sakstein:2017xjx,LIGOScientific:2018dkp}.
Indicatively, the inflationary phenomenology in the presence of the Gauss-Bonnet term has been presented in Ref.\cite{Oikonomou:2020sij,Odintsov:2020xji,Venikoudis:2021irr,Venikoudis:2021oee,Oikonomou:2021kql,Odintsov:2020zkl,Kanti:2015pda,Fomin:2020hfh,Pozdeeva:2020apf,Yi:2018gse,Wu:2017joj,Rashidi:2020wwg}. Before we proceed however, it is worth justifying the reasoning behind such constraint in the Gauss-Bonnet scalar coupling function in Eq. (\ref{ct}). The GW170817 event took place in the late-time era and currently there exists no indication that suggests that primordial tensor perturbations propagate through space time with the velocity of light. Hence, one could argue that such velocity could evolve dynamically and obtain a fixed value during the dark energy era. We select to work with a model described by a fixed value of the gravitational wave velocity only because the previous expression would describe primordial massive gravitons that need to become massless in late stages of the evolution of the Universe. Since we are unaware of the existence of such mechanism, we argue that the gravitational wave velocity had and 
always has a fixed value which in Natural units is $c_\mathcal{T}=1$, see Ref. \cite{Odintsov:2020sqy} for further implications of such constraint. In consequence, this constraint should impact the inflationary phenomenology compared to an unconstrained model. The result is that the effective degrees of freedom are decreased and one needs to specify only a single scalar coupling function, either the scalar potential $V(\phi)$ or the Gauss-Bonnet scalar coupling function $\xi(\phi)$. To showcase this explicitly, let us first incorporate the slow-roll conditions, in particular we assume that,
\begin{align}
\centering
\label{slowroll}
\dot H&\ll H^2&\frac{1}{2}\dot\phi^2&\ll V&\ddot\phi&\ll H\dot\phi.\
\end{align}
The condition $\ddot\phi\ll H\dot\phi$ is not only useful since it simplifies the continuity equation of the scalar field, it can also specify the time derivative of the scalar field from the constraint of the gravitational wave velocity. In particular, noting that the first and the second time derivatives of the scalar coupling function are $\dot\xi=\dot\phi\xi'$ and $\ddot\xi=\ddot\phi\xi'+\xi''\dot\phi^2$ respectively, it is obvious that the aforementioned constraint obtains the following form,
\begin{equation}
\centering
\label{xiconstraint}
\ddot\phi\xi'+\dot\phi^2\xi''=H\dot\phi\xi'\, ,
\end{equation}
and since $\ddot\phi\ll H\dot\phi$, it becomes apparent that the first time derivative of the scalar field can be written as,
\begin{equation}
\centering
\label{fdot}
\dot\phi\simeq H\frac{\xi'}{\xi''}.
\end{equation}
This expression is extracted by neglecting the irrelevant case for $\dot\phi=0$ and assuming that the Gauss-Bonnet scalar coupling function has a well-defined second derivative. As a result, by imposing the constraint of the velocity of primordial tensor perturbations, the degrees of freedom have decreased and now the continuity equation for the scalar field serves as a differential equation in order to extract the form of one scalar function while the other is appropriately specified. In fact, the equations of motion in the inflationary era under the slow-roll assumption obtain the following form (see Ref. \cite{Odintsov:2020mkz,Oikonomou:2020tct,Oikonomou:2020oil} for the constant roll case),
\begin{equation}
\centering
\label{motion1a}
H^2\simeq\frac{\kappa^2 V}{3}\, ,
\end{equation}

\begin{equation}
\centering
\label{motion2a}
\dot H\simeq-\frac{H^2}{2}\left(\frac{\kappa\xi'}{\xi''}\right)^2\, ,
\end{equation}

\begin{equation}
\centering
\label{motion3a}
V'+3H^2\frac{\xi'}{\xi''}+24\xi' H^4\simeq0.\
\end{equation}
In order to proceed with the cosmological dynamics during the inflationary era, we shall use the slow-roll indices where, 
based on the Ref.\cite{Hwang:2005hb} have the following form,
\begin{align}
\centering
\label{indices}
\epsilon_1&=-\frac{\dot H}{H^2},&\epsilon_2&=\frac{\ddot\phi}{H\dot\phi},&\epsilon_3&=\frac{\dot E}{2HE},&\epsilon_4&=\frac{\dot Q_t}{2HQ_t}.\
\end{align}
The first two slow-roll indices $\epsilon_1$, $\epsilon_2$ are described by simple forms, due to the minimal coupling between the Ricci scalar and the inflaton while, indices $\epsilon_3$ and $\epsilon_4$
have more complicated forms because they represent the impact of the Gauss-Bonnet gravitational term during inflation and specifically the index $\epsilon_4$ involves the significance of the Gauss-Bonnet scalar coupling function $\xi(\phi)$ relative to the Ricci scalar. In the following, we showcase the generalized expressions of the slow-roll indices for arbitrary scalar functions $V(\phi)$ and $\xi(\phi)$,
\begin{equation}
\centering
\label{e1general}
\epsilon_1 \simeq \frac{1}{2}\bigg(\frac{\kappa\xi'}{\xi''}\bigg)^2,
\end{equation}
\begin{equation}
\centering
\label{e2general}
\epsilon_2 \simeq 1-\epsilon_1-\frac{\xi '\xi''' }{ \xi ''^2},
\end{equation}
\begin{equation}
\centering
\label{e4general}
\epsilon_4 \simeq \frac{4 \kappa ^4 \xi '^2 \left(\xi ' \xi '' V'+V \left(2 \xi ''^2-\xi ^{'''} \xi '\right)\right)}{\xi ''^2 \left(8 \kappa ^4 V \xi '^2-3 \xi ''\right)},
\end{equation}
where we omit the index $\epsilon_3$ due to its quite complicated form.
In addition, according to Ref.\cite{Hwang:2005hb}, the auxiliary function $E(\phi)$ presented in index $\epsilon_3$
is given by the following expression,
\begin{equation}
\centering
\label{E}
E=\frac{1}{\kappa^2\dot \phi^2}\left(\dot \phi^2+\frac{3Q_a^2}{2Q_{t}}\right)\,
\end{equation}
where the function $Q_a$ is defined as $Q_a=-8\dot\xi H^2$.

In the following, we shall present under which conditions an inflationary model can be considered as viable based on the most recent Planck data in Ref.\cite{Planck:2018jri}. In the context of Einstein-Gauss-Bonnet theory the observational constraints, namely the scalar and tensor spectral index of primordial perturbations $n_\mathcal{S}$ and $n_\mathcal{T}$ respectively and the tensor-to-scalar ratio $r$ are given by the following expressions,
\begin{equation}
\centering
\label{scalarindex}
n_\mathcal{S} \simeq 1-2(2\epsilon_1+\epsilon_2+\epsilon_3)\, ,
\end{equation}

\begin{equation}
\centering
\label{tensorindex}
n_\mathcal{T} \simeq-2(\epsilon_1+\epsilon_4)\, ,
\end{equation}

\begin{equation}
\centering
\label{ratio}
r \simeq 16\left|\left(\epsilon_1-\frac{\kappa^2Q_e}{4H}\right)\frac{c_A^3}{\kappa^2Q_t}\right|\, ,
\end{equation}
where $c_A$ indicates the field propagation velocity given by the expression,
\begin{equation}
\centering
\label{soundwave}
c_A^2=1+\frac{Q_aQ_e}{2Q_{t}\dot \phi^2+3Q_a^2},\,
\end{equation}
where $Q_e=-32\dot \xi \dot H$ is an additional auxiliary function. As it is expected, the equations of the observational indices are strongly depended from the slow-roll parameters. 

According to the most recent Planck data, the scalar spectral index of primordial perturbations takes the numerical value
$n_\mathcal{S}=0.9649\pm 0.0042 $
with $67\%$ C.L while the tensor-to-scalar-ratio r must be constrained in the spectrum $r<0.064$ with $95\%$ C.L. Concerning the tensor spectral index, due to the fact that B-modes have yet to be spotted in the CMB, there is no bound on its value hence, it can be either red-tilted with negative value or blue-tilted with $n_\mathcal{T}>0$ \cite{Oikonomou:2021kql}. At this point of our analysis, we demonstrate the formalism which is used in order to evaluate the observational indices during the first horizon crossing. For convenience, let us use the initial and final values of the scalar field instead of wavenumbers. Firstly, we shall determine the final value of the inflaton by setting the first slow-roll index (\ref{e1general}) equal to unity. After the evaluation of the value $\phi_f$ we shall use the e-folding number equation, in order to determine the value of the scalar field during the first horizon crossing $\phi_i$. Hence, based on the equation $N=\int_{t_k}^{t_f}{Hdt}=\int_{\phi_k}^{\phi_f}{\frac{H}{\dot\phi}d\phi}$, where the difference $t_f-t_i$ indicates the duration of the
inflation and according to the equation (\ref{fdot}), the initial value $\phi_k$ can be determined from the following equation,
\begin{equation}
\centering
\label{efoldsk}
N_k=\int_{\phi_k}^{\phi_f}{\mathrm{d}\phi\frac{\xi''}{\xi'}}\, , 
\end{equation}
where the duration of the inflationary era ranges in the interval $N_k \sim 50-60$ e-folding.
As it is demonstrated above, the e-folding number equation in Gauss-Bonnet theories does not involve the scalar coupling function $\xi(\phi)$ per se but it is strongly related with the ratio of their derivatives. Thus, it is reasonable to select convenient coupling functions that greatly simplify such expression. The choice of an exponential scalar coupling seems to fail as the first slow-roll index becomes $\phi$ independent and depending on the exponent of $\xi$, it can describe eternal or no inflation at all. There exist two choices that seem to simplify the aforementioned ratio \cite{Odintsov:2020sqy}, namely the power-law model and the error function model. For simplicity we shall limit our work to only one of them, in particular the error function model in the subsequent sections.

\section{Numerical Solutions of the Models}
In the last section of this paper we
present the numerical results, which arise from the investigation of the phenomenology of two cosmological models of interest. In the first model, we consider that the Gauss-Bonnet scalar coupling function has the specific form of the error function while, in the second case the scalar potential is given by the Woods-Saxon potential.
For both models the scalar potential and the scalar Gauss-Bonnet coupling function are interconnected, which implies that knowledge of one scalar function allows us to completely specify the other from the continuity equation (\ref{motion3a}). The initial point of the analysis is the first horizon crossing. After designating the free parameters
of the models by achieving compatibility for the
observational indices with the latest Planck 2018 collaboration
data \cite{Planck:2018jri}, we focus on the reheating era. In particular, the duration of reheating in terms of the e-folding number and the reheating temperature are computed based on the equations (\ref{reefolds}) and (\ref{Tre}). For the first model we examine whether the stiff matter can be considered as a viable option for such era while, for the Woods-Saxon model we propose two different scenarios for the reheating era depending from the potential amplitude and the EoS parameter.

\subsection{Inflationary Dynamics}
Before proceeding with the numerical results, it is worth mentioning a key feature of the theory. Since the Gauss-Bonnet scalar coupling function is constrained due to the realisation that tensor perturbations propagate with the velocity of light, the continuity equation of the scalar field, as mentioned before, serves as a differential equation, either first order for the scalar potential or second order for the Gauss-Bonnet scalar coupling function. For the sake of generality, we shall cover both approaches in the following models. In fact, for the first model the continuity equation shall be treated as a first order differential equation with respect to the canonical potential while, on the second case it shall be treated as a second order differential equation with respect to the Gauss-Bonnet scalar coupling function. While a first order differential equation can  be solved, in the second case the form of the Gauss-Bonnet scalar coupling function is not guaranteed to be simple however the ratio $\frac{\xi'}{\xi''}$, which is of paramount importance, is simplified.

\subsubsection{Error function}
For the first model we shall assume that the Gauss-Bonnet scalar coupling function is in fact specified by the following expression,

\begin{equation}
\centering
\label{xi1}
\xi(\phi)=Erf(\kappa\phi)\, ,
\end{equation}
which was introduced in Ref.\cite{Odintsov:2020sqy}. This choice simplifies the overall phenomenology since the ratio $\xi'/\xi''$ turns out to be linear with respect to the scalar field therefore the first slow-roll index (\ref{e1general}) obtains a quite convenient form as we shall showcase subsequently. Now by treating the continuity equation as a first order differential equation with respect to the scalar potential, one finds that for the error function model,

\begin{equation}
\centering
\label{v1}
V(\phi)=\frac{3 \sqrt{\pi } \phi ^{\frac{1}{2}} \left( \kappa ^2 \phi ^2\right)^{\frac{3}{4}}}{3 \sqrt{\pi } V_1 \left( \kappa ^2 \phi ^2\right)^{\frac{3}{4}}-8   \kappa ^5\phi ^{\frac{3}{2}} \Gamma \left(\frac{3}{4},\kappa ^2 \phi ^2\right)}
\end{equation}
where $V_1$ is the integration constant with mass dimensions of eV$^{-7/2}$. As showcased, the potential has a nontrivial form and according to Ref. \cite{Odintsov:2020sqy}, the integration constant is required since a viable inflationary era is obtained only by fine tuning such parameter. The first two slow-roll indices are given by the following expressions,

\begin{equation}
\centering
\label{index1}
\epsilon_1=\frac{1}{8 \kappa ^2 \phi ^2}\, ,
\end{equation}

\begin{equation}
\centering
\label{index2}
\epsilon_2=3\epsilon_1\, ,
\end{equation}
whereas the rest slow-roll indices are neglected due to their perplexity however their contribution is significant and affect the spectral indices strongly. In consequence, by evaluating the scalar field at the end of the inflationary era one extracts the following solution,

\begin{equation}
\centering
\label{phif}
\phi_f=\pm\frac{1}{2 \sqrt{2} \kappa }.\
\end{equation}
For consistency, we shall keep the negative value since it corresponds to a positively defined scalar potential with respect to the free parameters which shall be used. In addition, according to the e-folding equation (\ref{efoldsk}) the value of the scalar field during the first horizon crossing becomes

\begin{figure}[h!]
\centering
\includegraphics[width=16pc]{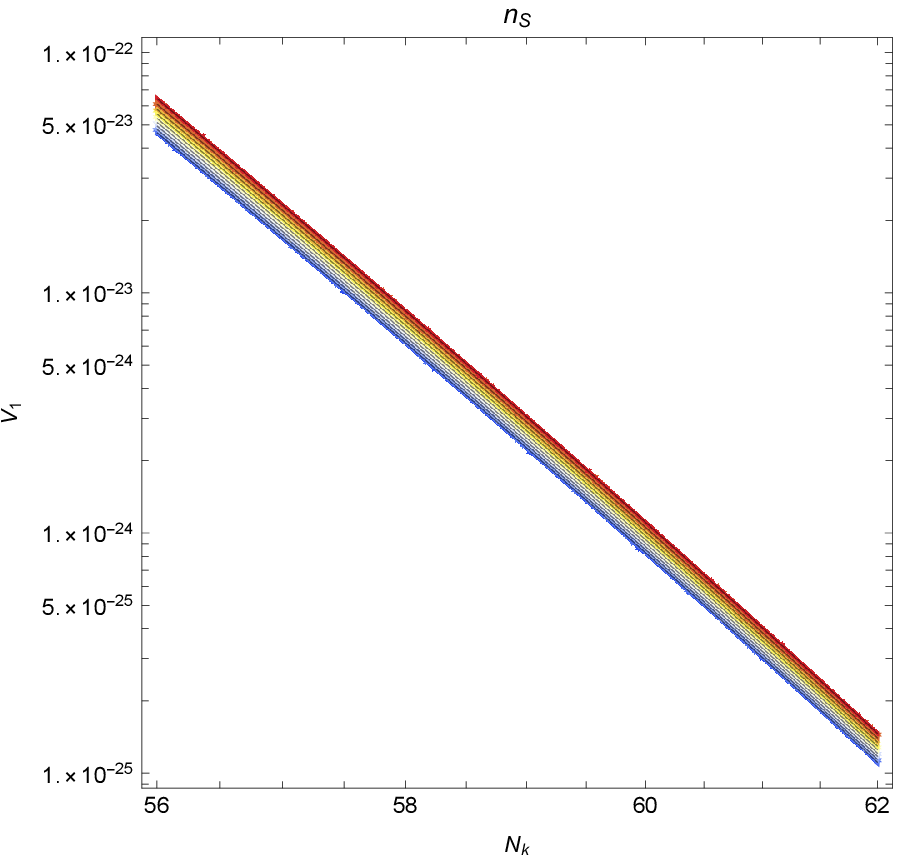}
\includegraphics[width=3pc]{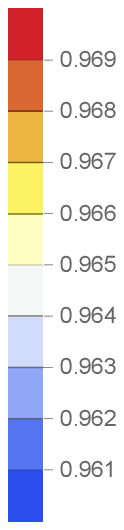}
\includegraphics[width=16pc]{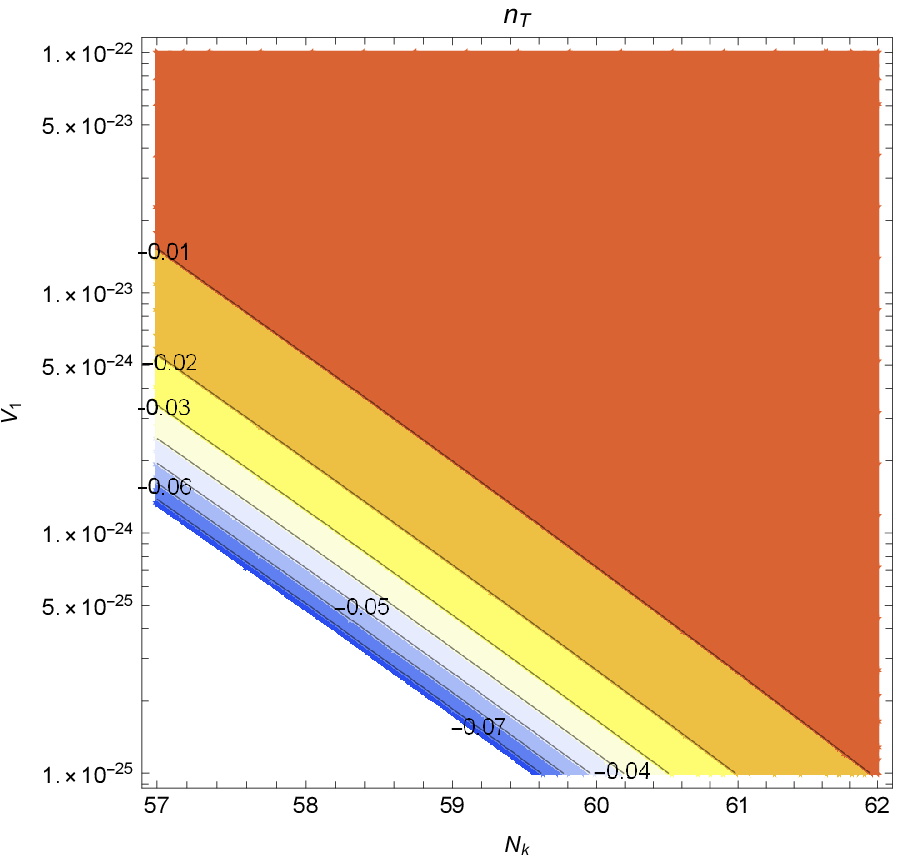}
\includegraphics[width=2.8pc]{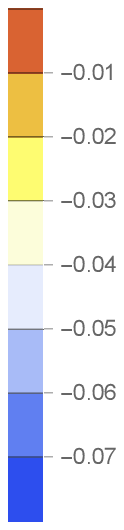}
\caption{Scalar spectral index and tensor spectral index for the error function model. Parameter $V_1$ of the scalar potential is given
in logarithmic scale. The tensor-to-scalar ratio was neglected since it is affected mostly by the e-folding number $N_k$.}
\label{spectralerf}
\end{figure}

\begin{equation}
\centering
\label{phik}
\phi_k=\pm\frac{1}{\kappa}\sqrt{\frac{1}{8}+N_k}\, ,
\end{equation}
where, hereafter only the positive value shall be considered. Let us proceed with the designation of the free parameters. Assuming that $(N_k, V_1)=(60, 10^{-24})$ one finds that the observed indices are in agreement with the observational constraints as the scalar spectral index is equal to $n_\mathcal{S}=0.966686$, the tensor to scalar ratio is $r=0.0332632$ and the prediction for the tensor spectral index is $n_\mathcal{T}=-0.00836797$. It is worth mentioning that due to the string corrections, the relation $r\simeq-8n_\mathcal{T}$ which appears in the case of the canonical scalar field is not respected. As mentioned before, parameter $V_1$ must be fine tuned in order to obtain a viable value for the scalar spectral index otherwise, viability is not obtained. This can easily be inferred from Fig.\ref{spectralerf}.
The initial value of the inflaton during the first horizon crossing is $\phi_k\simeq 7.75$ and its final value when the inflationary era ceases is $\phi_f\simeq -0.35$, which means that the scalar field decreases in the inflationary epoch. In the following, we present the numerical values of the slow-roll parameters during the first horizon crossing. The slow-roll indices $\epsilon_1$ and $\epsilon_4$ are almost equal $\epsilon_1\simeq\epsilon_4\simeq 0.002$ while $\epsilon_2\simeq\epsilon_3\simeq 0.0062$ thus, the slow-roll parameters are all in the same order of magnitude. In addition, we stress that the field propagation velocity is almost equal to unity respecting the causality.

In order to ascertain the validity of the model we examine whether our approximations during the inflationary era are accurate. In the following, we present the values of the slow-roll conditions in order of magnitude during the first horizon crossing. Specifically, 
$\dot H \sim \mathcal{O}(10^{21})$ and $ H^2 \sim \mathcal{O}(10^{23})$. In addition, the numerical value of the 
the kinetic term of the inflaton $\frac{1}{2}\dot\phi^2\sim\mathcal{O}(10^{21})$ is minor compared to the scalar potential which is $V(\phi)\sim \mathcal{O}(10^{24})$ indicating that the condition $\frac{1}{2}\dot\phi^2\ll V(\phi)$ holds true and finally the approximation $\ddot \phi \ll 3H\dot\phi$ is satisfied since $\ddot \phi \sim \mathcal{O}(10^{20})$ and
$3 H\dot \phi\sim\mathcal{O}(10^{23})$. Hence, the slow-roll approximations are indeed valid thus satisfying the reason they were neglected in the first place.

Moreover, we evaluate the Gauss-Bonnet string corrections during the first horizon crossing. The term $16\dot \xi H \dot H \sim \mathcal{O}(10^{19})$ is minor compared to the kinetic term and the term $24\dot \xi H^3\sim\mathcal{O}(10^{22})$ can be considered as negligible compared to the scalar potential $V(\phi)$. The numerical results indicate that the model is validate.

\begin{figure}[h!]
\centering
\includegraphics[width=18pc]{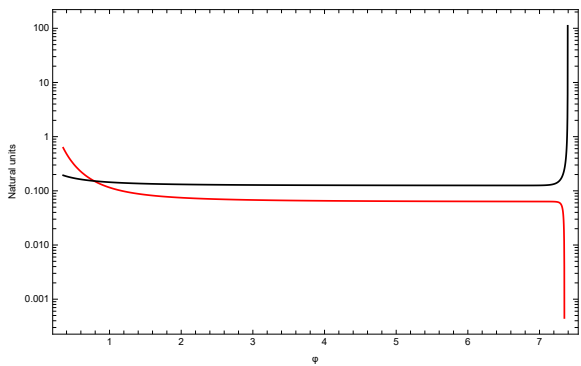}
\caption{Logarithmic representation of the second time derivative of Gauss-Bonnet coupling function $\ddot \xi$ (red) and the product of $H\dot{\xi
}$ (black) during the inflationary era.}
\label{constrainterf}
\end{figure}

Last but not least, given that the whole phenomenology of model was examined based on the constraint of the velocity of the Gravitational Wave 170817, the validity of the Eq.(\ref{X1}) should be examined in detail. Fig.\ref{constrainterf} illustrates that in the inflationary epoch the condition 
$\ddot \xi=H\dot \xi$ holds true, at least in order of magnitude. In particular, at the beginning of the inflation the numerical values of the physical quantities are $\ddot \xi=-0.000528443$ and $H\dot \xi=-0.000524076$ by indicating that the approximations that were set have exceptional accuracy. However, at the last stage of the inflationary era, in this tiny time interval before it starts the reheating of the Universe, the constraint equation is satisfied in order of magnitude since $\ddot \xi
\sim \mathcal{O}(10^{-1})$ and $H \dot \xi \sim \mathcal{O}(10^{-1})$. Overall, it becomes clear that the approximation is indeed valid and thus one can make use of $\ddot\xi$ and $H\dot\xi$ interchangeably.

\subsubsection{Woods-Saxon}
Let us now proceed with the second model. For the case at hand, the scalar potential is given by a Woods-Saxon form which is inspired from nuclear physics. The inflationary phenomenology and the reheating process has been presented for the Woods-Saxon canonical scalar field model in Ref. \cite{Oikonomou:2019gjj}. In this case the potential reads,
\begin{equation}
\centering
\label{WSpotential}
V(\phi)=\frac{\Lambda^4}{1+be^{c \kappa \phi}},
\end{equation}
where $\Lambda$, $b$ and $c$ are free parameters of the theory with dimensions eV for the first while, $b$ and $c$ are dimensionless constants.
Hereafter, in order to reduce the free parameters of the theory we consider that $b$ is equal to unity. Even though the gravitational action (\ref{action}) involves two scalar functions, as mentioned before, only one them is free due to the constraint in the propagation velocity of primordial tensor perturbations. Therefore by designating the scalar potential, the Gauss-Bonnet coupling function is extracted from the continuity equation (\ref{motion3a}) and has the following form,
\begin{equation}
\centering
\label{GBcoupling}
\xi(\phi)=\lambda\int^{\kappa\phi}e^{\frac{-e^{-c \kappa x}+c\kappa x}{c^2}}dx,
\end{equation}
where $x$ represents an auxiliary variable of integration and $\lambda$ is a dimensionless constant which shall be considered as equal to unity for simplicity. The slow-roll parameters of the theory are then given by the following equations,
\begin{equation}
\centering
\label{epsilon1WS}
\epsilon_1 \simeq \frac{c^2}{2 \left(e^{-c \kappa  \phi }+1\right)^2},
\end{equation}

\begin{equation}
\centering
\label{epsilon2WS}
\epsilon_2 \simeq -\frac{c^2 e^{c \kappa  \phi } \left(e^{c \kappa  \phi }-2\right)}{2 \left(e^{c \kappa  \phi }+1\right)^2},
\end{equation}

\begin{equation}
\centering
\label{epsilon4WS}
\epsilon_4 \simeq \frac{4 c \kappa ^3 \Lambda ^4 e^{\frac{\kappa \phi(c^2+1)}{c}} \left(\left(c^2-1\right) e^{2 c \kappa  \varphi }-\left(c^2+2\right) e^{c \kappa  \phi }-1\right)}{3 e^{\frac{e^{-c \kappa  \phi }}{c^2}} \left(e^{c \kappa  \phi }+1\right)^4-8 c \kappa ^3 \Lambda ^4 e^{\frac{\kappa \phi(c^2+1)}{c}} \left(e^{c \kappa  \phi }+1\right)^2},
\end{equation}
where index $\epsilon_3$ has been omitted due to its complicated form. The final value of the scalar field which signifies the end of the inflationary era is extracted by setting the first slow-roll index in equation
(\ref{epsilon1WS}) equal to unity. In this case we obtain,
\begin{equation}
\centering
\label{phifinal}
\phi_f\simeq \frac{\log \left(\frac{\sqrt{2}(c+\sqrt{2})}{c^2-2}\right)}{c \kappa }.
\end{equation}
From the e-folding equation (\ref{efoldsk}) one can extract the initial value of the inflaton where in this model has a complicated form and is thus omitted.

\begin{figure}[h!]
\centering
\includegraphics[width=18pc]{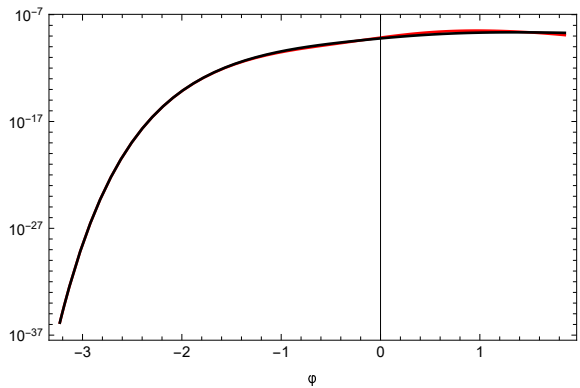}
\caption{Graphic illustration of the second time derivative of Gauss-Bonnet coupling function $\ddot \xi$ (red) and the product of $H\dot{\xi
}$ (black) in logarithmic scale during inflation.}
\label{constraintWS}
\end{figure}

At this stage, let us specify the free parameters of the theory in order to examine whether the particular model can be viable based on the most recent Planck data \cite{Planck:2018jri}. We assume the following two set of numerical values ($N_k$, $\Lambda$, c)=(60, 0.01, 1.5) for the free parameters in reduced Planck units where $\kappa^2=1$. According to the aforementioned
numerical values, the scalar spectral index of primordial perturbations and the tensor-to-scalar ratio are in agreement with the observational constraints since $n_\mathcal{S}=0.965093$ and $r=0.00109162$.
The tensor spectral index takes the numerical value $n_\mathcal{T}=-0.000136452$ indicating a slightly red-tilted spectrum as usual in Gauss-Bonnet gravitational theories. The inflaton takes the initial value $\phi_k=-3.23161$ and increases during the inflationary era with time since it reaches the final value $\phi_f=1.86831$ when the inflation ceases. In addition, the field propagation velocity is equal to unity $c_A^2=1$ indicating the absence of ghosts instabilities. We also present the numerical values of the slow-roll indices during the first horizon crossing in order of magnitude. More specifically, $\epsilon_1\sim \mathcal{O}(10^{-5})$, $\epsilon_2 \sim \mathcal{O}(10^{-2})$, $\epsilon_3\sim \mathcal{O}(10^{-67})$ and $\epsilon_4\sim \mathcal{O}(10^{-36})$. The slow-roll indices $\epsilon_3-\epsilon_4$ which involve the impact of Gauss-Bonnet string-corrections during the inflationary dynamics are effectively zero. Since their numerical values are infinitesimal compared with the indices $\epsilon_1$ and $\epsilon_2$, once can easily realize that string corrections do not influence the inflationary phenomenology of the Woods-Saxon model. This is due to the effective absence of the term $24\xi'H^4$ in the continuity equation which is quite insignificant and does not contribute, to the point where keeping such term does not seem to alter the numerical value of $\xi$ which is the reason it was omitted in the first place.

In the following we shall examine whether the approximations which were made in order to simplify the equations of motions hold true.
Firstly, we shall evaluate the
slow-roll conditions (\ref{slowroll}) during the first horizon crossing. As expected
$\dot H \sim \mathcal{O}(10^{-13})$ and $H^2 \sim \mathcal{O}(10^{-9})$ hence 
the slow-roll conditions are valid. Furthermore, the numerical contribution of the kinetic term of scalar field is negligible compared to the Woods-Saxon potential as $\frac{1}{2}\dot \phi^2 \sim \mathcal{O}(10^{-13})$ and $V \sim \mathcal{O}(10^{-9})$ with 
the condition $\frac{1}{2}\dot \phi^2\ll V(\phi)$ being satisfied. Last but not least, $\ddot \phi \sim \mathcal{O}(10^{-13})$ and 3$H \dot \phi \sim \mathcal{O}(10^{-10})$ concluding that the set of approximations in Eq.(\ref{slowroll}) is accurate.
Moreover, in Fig.\ref{constraintWS} we present the excellent accuracy of the condition $\ddot \xi=H\dot \xi$ in the inflationary epoch. In the first horizon crossing the numerical values of the aforementioned quantities are $\ddot \xi=1.16408\cdot 10^{-36}$
and $H\dot \xi=1.14427 \cdot 10^{-36}$ in Natural units while, an infinitesimal deviation, which is insignificant, between the two quantities arises when the inflation ceases. Hence, as was the case with the previous model, the constraint imposed on the primordial gravitational wave velocity is proven to be true on a numerical level.

\begin{figure}[h!]
\centering
\includegraphics[width=16pc]{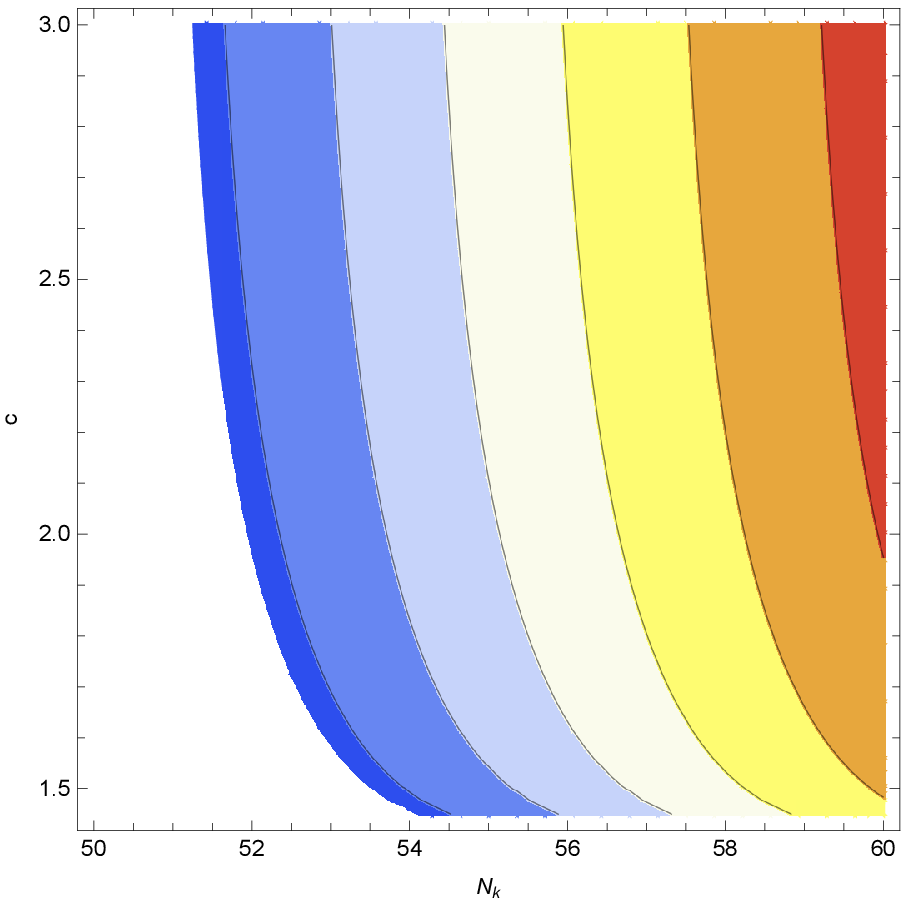}
\includegraphics[width=3pc]{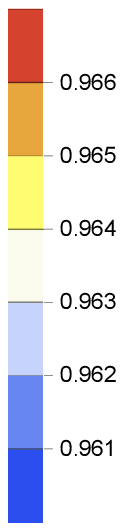}
\includegraphics[width=16pc]{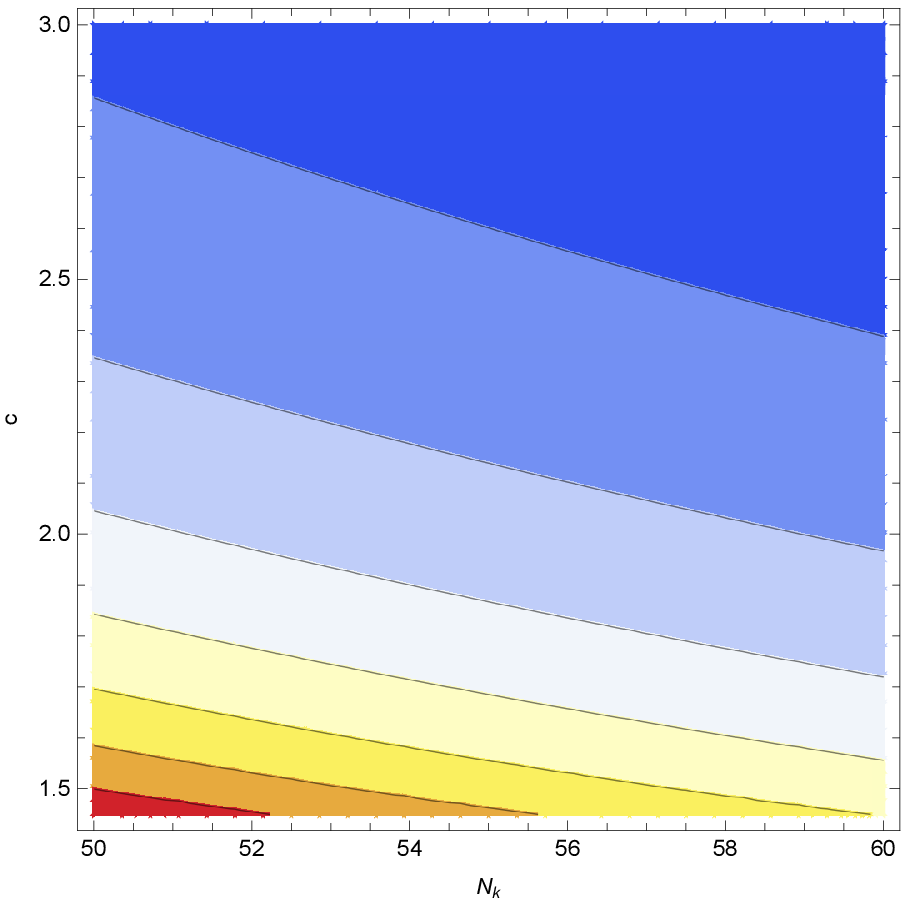}
\includegraphics[width=3.3pc]{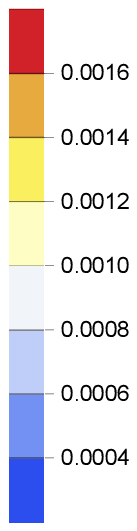}
\caption{Contour plots of scalar spectral index of primordial perturbations and the tensor-to-scalar ratio index depending on the e-folding number $N_k$ and the free parameter $c$ for the Woods-Saxon model.}
\label{plot1}
\end{figure}

All that remains is the numerical evaluation of Gauss-Bonnet corrections
during the first horizon crossing. More specifically, the term $24\xi' H^4 \sim \mathcal{O}(10^{-42})$ can be considered negligible compared to the derivative of scalar potential $V'\sim \mathcal{O}(10^{-10})$, as a result the continuity equation of the scalar field 
is satisfied. Furthermore, $16\dot \xi H \dot H\sim  \mathcal{O}(10^{-48})\ll\frac{1}{2}\dot \phi^2$ and $24\dot\xi H^3\sim\mathcal{O}(10^{-44})$ is also minor compared with $V\simeq 1$. Given that our approximations hold true the model can be considered as viable. The reason why string corrections are so subleading is due to the double exponential form of $\xi(\phi)$ which, for a negative $\phi_k$, vanishes.

Finally, in Fig.\ref{plot1} we demonstrate how the scalar spectral index of primordial perturbations and the tensor-to-scalar-ratio are affected by the free parameters of the theory. As shown, both indices are affected by the parameters but the tensor-to-scalar-ratio has a stronger dependence on them. As a last note we stress that a viable inflationary era can also be achieved using different values of the free parameter $\Lambda$. More specifically,
the case where $\Lambda=9.1\cdot 10^{-6}$,
also leads to validate results since, the observational indices take accepted values according the latest Planck data. In the following, we demonstrate the cosmological behaviour in the reheating era for the two aforementioned values of the potential amplitude $\Lambda$.

\subsection{Reheating Era}
In this subsection we showcase the numerical results which arise from the transition of the inflationary era to the reheating era of the primordial Universe. During the reheating phase, the existed matter content will be thermalized due to the energy density and subsequently will affect the primordial Big Bang Nucleosynthesis. In order to achieve smooth transition between the two eras we focus on the final value of the inflaton. Firstly, we solve Eq. (\ref{X2}) algebraically for both models with respect to $X(\phi_f)$ in order to evaluate the value of this function when the inflation ceases. The aforementioned function is the only object which depends on the Gauss-Bonnet scalar coupling function and differs between string-inspired gravity and the canonical scalar field case. In the following, we present the numerical results of the reheating era for the two models considering two possible values of the free parameter $\Lambda$ on the Woods-Saxon model.

\subsubsection{Reheating era for the Error function model}
In order to extract the numerical value of the reheating temperature at the beginning of the radiation domination era and the duration of the reheating era, one must first compute parameter $X$ for $\phi=\phi_f$. According to the previous formalism and for the same set of parameters for the Error function model, one finds that $X(\phi_f)=-1.4748$ in Natural units which, as showcased is greater than unity but only in absolute value in order to obtain a proper description for the e-folding number. Note that the sign is in agreement with the evolution of the scalar field since, the value of the scalar field during the inflationary era was chosen such that it decreases.

\begin{figure}[h!]
\centering
\includegraphics[width=19pc]{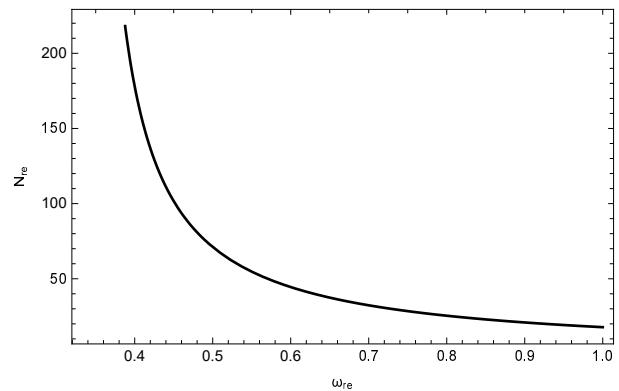}
\caption{Duration of reheating as a function of $\omega_{re}$ in the area of EoS parameter (1/3, 1] for the choice of error function as Gauss-Bonnet scalar coupling function.}
\label{efoldserf}
\end{figure}

\begin{figure}[h!]
\centering
\includegraphics[width=19pc]{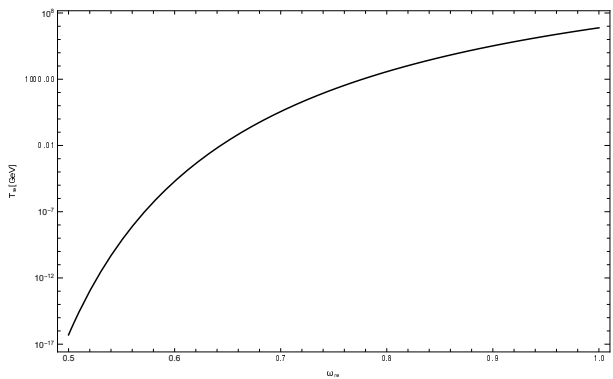}
\caption{Reheating temperature versus the effective EoS in the area of EoS parameter [1/2, 1].}
\label{temperf}
\end{figure}

Now, assuming that the effective EoS parameter during the reheating era is equal to unity, that is $\omega_{re}=1$, the duration of the inflationary era is $N_k=60$ and the Hubble rate during the first horizon crossing is $H_k\simeq 1.7 \cdot 10^{-6}$ in Natural units, one finds that the e-folding number and the reheating temperature are equal to
$N_{re}\simeq17$ and
$T_{re}=7.6\cdot10^6$ GeV from equations (\ref{reefolds}-\ref{Tre}) respectively. It is worth mentioning that the duration of the reheating is quite large, however it depends strongly on the effective EoS parameter for such era and decreases with the increase of $\omega_{re}$, as shown in Fig\ref{efoldserf}. In contrast, the reheating temperature increases with the increase of the EoS parameter according to Fig.\ref{temperf}. Specifically, if one considers that $\omega_{re}$ ranges in the vicinity of that of radiation the reheating temperature is extremely small, to such a degree that subsequent cosmological eras cannot be described as $T_{re}$ is below the MeV scale. As a general statement, it can easily be inferred that small values of $\omega_{re}$ suggest longer duration however smaller reheating temperature, thus making the choice of stiff matter the ideal scenario. We also highlight that the model is incapable of producing a viable reheating era for $\omega_{re}<1/3$ since, the number of the e-folds becomes negative.

The case of stiff matter represents the maximum allowed EoS which respects causality and described in the context of the kinetic regime. Specifically, the energy density is scaling as $\rho\sim \alpha(t)^{-6}$ with the scale factor of the Universe by behaving as $\alpha(t) \sim t^{1/3}$ by indicating that the energy density scales twice as fast as ordinary dust.

\subsubsection{Reheating era for the Woods-Saxon model with $\Lambda=0.01$ in Natural units}
We proceed with the numerical results of the reheating era for the second model which involves the scalar potential of the Woods-Saxon form.
Similar to the previous case, we extract the value of $X(\phi_f)$ from the equation (\ref{X2}) with the potential amplitude being $\Lambda=0.01$. We get that when the inflationary era ceases, the numerical value of the function is $X(\phi_f)\simeq - 10^{-8}$ in Natural units, which is in consistency with our predictions given that the denominator in Eq. (\ref{X2}) takes positive value.
A comparison between the auxiliary functions $X(\phi_f)$ of both models indicates that string corrections do not seem to influence strongly the overall phenomenology in the Woods-Saxon model.

At this point we can proceed with the e-folding number and the reheating temperature of this era, which can be extracted from equations (\ref{reefolds}) and (\ref{Tre}) respectively. It is assumed that the potential energy which originates from the Woods-Saxon potential decays into particles which fill the Universe. In this case the corresponding EoS parameter takes the value $\omega_{re}=1$ for stiff matter. Given that the e-folding number during the inflation is $N_k=60$ and the value of the Hubble rate during the first horizon crossing is $H_k\simeq 3.2 \cdot 10^{-6}$ in Natural units, we obtain that the e-folding number during the reheating phase is $N_{re}\simeq 11$ and the reheating temperature is $T_{re}=7.7 \cdot 10^{8}$ GeV.
As indicated from the Planck temperature which is $T_{Pl}\sim\mathcal{O}(10^{19})$ Gev, the temperature of the Universe dropped significantly
due to the quasi-exponential expansion during the inflationary era.
\begin{figure}[h!]
\centering
\includegraphics[width=18pc]{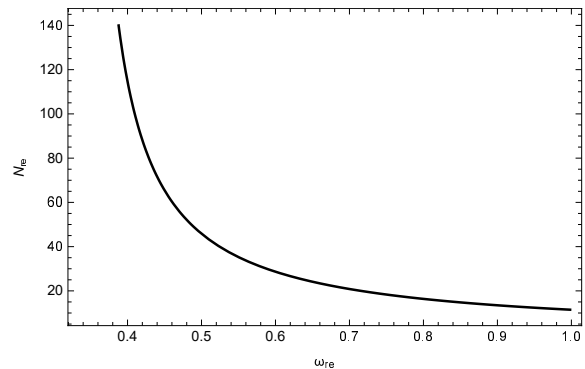}
\caption{Numerical solution of the e-folding number during the reheating era with respect to the equation of state parameter in the area (1/3,1]. As depicted, the e-folding number drops as the EoS parameter increases and reaches its minimum value for the case of stiff matter.}
\label{plot2}
\end{figure}

\begin{figure}[h!]
\centering
\includegraphics[width=18pc]{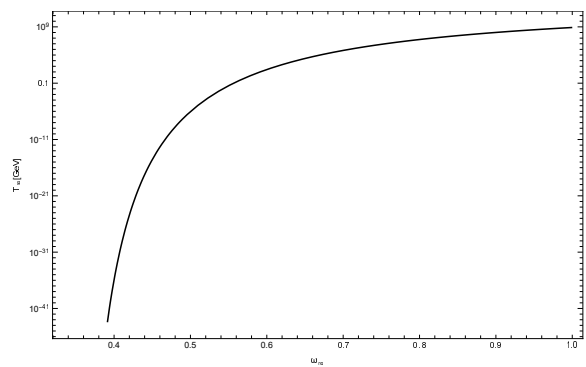}
\caption{Logarithmic plot of the numerical solution of the reheating temperature as function of the EoS in the area (1/3,1]. The temperature values are given in Gev.}
\label{plot3}
\end{figure}

In Fig.\ref{plot2} we demonstrate the behaviour of the e-folding number with respect to the EoS in the area $(1/3,1]$. The e-folding number takes unacceptable values for the case of radiation domination since it is predicted that the duration of the reheating era is approximately three times bigger than the inflationary era meaning that the Universe needs an important amount of time in order to be thermalized. In addition, the number of e-folds takes negative values if one consider that $-1/3<\omega_{re}<1/3$. whereas the stiff matter scenario leads to viable results. Moreover, in Fig.\ref{plot3} the behaviour of the reheating temperature is presented as a function of the EoS parameter. As depicted, according to this scenario the Universe cannot be reheated if it involves relativistic matter.

\subsubsection{Reheating era for the Woods-Saxon model with $\Lambda=9.1 \cdot 10^{-6}$ in Natural units}

In order to examine the reheating era for this infinitesimal value of the potential amplitude $\Lambda$, we follow the same process as presented before. The algebraic value of the function $X(\phi_f)$ when the inflation ends is $X(\phi_f)\simeq -7\cdot 10^{-17}$. As demonstrated previously, the quantity $(1-\kappa^2 X)$ remains positive and the energy density also takes positive values. In this case we consider an alternative scenario where the Universe is described by the EoS parameter with $\omega_{re}=1/5$. This value is close to the case of the relativistic matter, which corresponds to $\omega=1/3$. We proceed with the reheating phenomenology by evaluating the 
reheating temperature and the number of e-folding. The numerical results are $T_{re}\simeq 2.2 \cdot 10^{8}$ GeV and $N_{re}\simeq 13$.

\begin{figure}[h!]
\centering
\includegraphics[width=18pc]{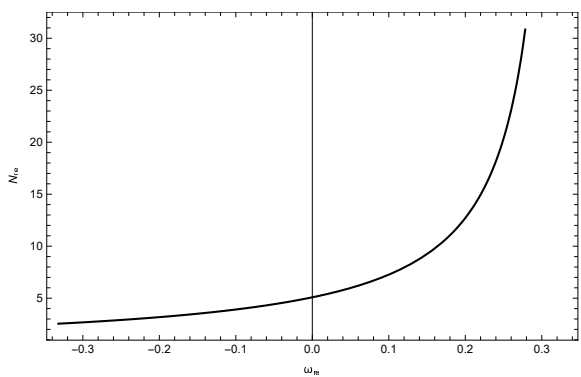}
\caption{Plot of the e-folding number with respect to the EoS parameter in the area of $\omega_{re}$ (-1/3,1/3) for the case of $\Lambda=9.1\cdot 10^{-6}$ in Natural units. As shown, the duration of reheating increases for positive values of the EoS parameter and especially close to the case of radiation. }
\label{plot4}
\end{figure}

\begin{figure}[h!]
\centering
\includegraphics[width=19pc]{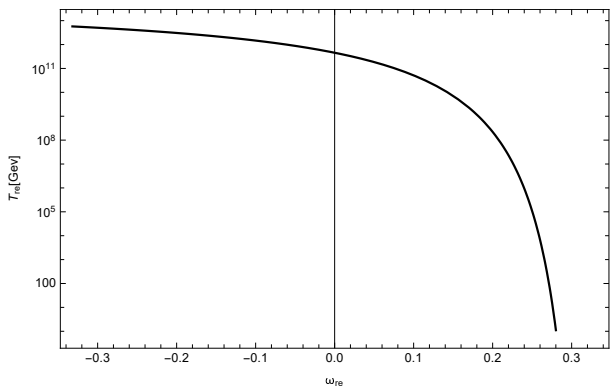}
\caption{Logarithmic plot of the reheating temperature as function of the EoS parameter in the area of $\omega_{re}$ (-1/3, 1/3). In this case, larger values of the EoS result in a smaller value for the reheating parameter. The plot corresponds to $\Lambda=9.1\cdot 10^{-6}$ in Natural units.}
\label{plot5}
\end{figure}

In Figures \ref{plot4} and \ref{plot5} the behaviour of the e-folding number and the reheating temperature are shown as functions of the EoS parameter. As the number of e-folds becomes negative in the vicinity $\omega_{re}(1/3,1]$, the model is incapable of producing a viable reheating era, if the $\omega_{re}$ ranges on the aforementioned vicinity.

Lastly, comparing the two possible scenarios, it is obvious that the amplitude of the Woods-Saxon potential $\Lambda$ dictates the allowed numerical values of the EoS parameter. A large value of the scalar potential at the final stage of the inflationary era is connected to a large EoS parameter in order to obtain a viable reheating era whereas small values of the scalar potential are consistent with small values of the EoS, even close to the case of radiation.

\section{Conclusions}
In this work we presented an alternative formalism for the description of the reheating era of primordial Universe in the context of Einstein-Gauss-Bonnet gravitational theory. For consistency, we investigated firstly the inflationary era for two models of interest. In the first we considered that the Gauss-Bonnet scalar coupling function had the form of an error function and in the second that the scalar field is described by a Woods-Saxon potential. The observational constraints, namely the spectral index of primordial perturbations and the tensor to scalar ratio, were in agreement with the most recent Planck data. In order to describe a smooth transition to the thermalization process, we evaluated all the involved physical quantities at the last stages of the inflationary era. Then, for both models at hand the numerical values of the reheating temperature and the e-folding number where determined.
For the error function model we considered that the Universe involved stiff matter where $\omega_{re}=1$ while for the second model we assumed the stiff matter scenario and an EoS parameter quite close to the radiation dominated era with $\omega_{re}=\frac{1}{5}$. The distinction between the two cases appears by allowing different values of the potential amplitudes. In both cases it was shown that a viable reheating era can indeed be obtained for the same set of values for the free parameters that produce a viable inflationary era and the dependence of the reheating temperature and the duration of reheating were showcased explicitly. As demonstrated, not all values for the effective EoS are acceptable due to the fact that the reheating temperature may be lower than the MeV scale, which is necessary in order to observe nucleosynthesis subsequently, or the e-folding number was ill defined. Therefore, the constraint imposed on the velocity of primordial gravitational waves simplifies the overall phenomenology to quite an extend, to the point where the procedure is quite similar to the canonical scalar field, and manages to produce a viable reheating era. An interesting question that arises is whether a similar formalism applies to other string inspired terms when compatibility with the GW170817 event is assumed to hold in the early Universe as well. In addition, other string inspired theories which include the Chern-Simons parity violating term can be used in order to describe the reheating era. We hope to address these in future works.\\

\section{ACKNOWLEDGMENTS}
The authors would like to express their gratitude towards Dr. V.K. Oikonomou for his comments and suggestions on the Gauss-Bonnet gravitational theories, that improved the context and the quality of the paper.


\begin{thebibliography}{99}




\bibitem{Guth:1980zm}
A.~H.~Guth,
Phys. Rev. D \textbf{23} (1981), 347-356
doi:10.1103/PhysRevD.23.347


\bibitem{Oikonomou:2017bjx}
V.~K.~Oikonomou,
Mod. Phys. Lett. A \textbf{32} (2017) no.33, 1750172
doi:10.1142/S0217732317501723
[arXiv:1706.00507 [gr-qc]].


\bibitem{Mijic:1986iv}
M.~B.~Mijic, M.~S.~Morris and W.~M.~Suen,
Phys. Rev. D \textbf{34} (1986), 2934
doi:10.1103/PhysRevD.34.2934








\bibitem{Cook:2015vqa}
J.~L.~Cook, E.~Dimastrogiovanni, D.~A.~Easson and L.~M.~Krauss,
JCAP \textbf{04} (2015), 047
doi:10.1088/1475-7516/2015/04/047
[arXiv:1502.04673 [astro-ph.CO]].




\bibitem{Mathew:2020nqo}
A.~Mathew and M.~K.~Nandy,
[arXiv:2012.13960 [gr-qc]].


\bibitem{vandeBruck:2016xvt}
C.~van de Bruck, K.~Dimopoulos and C.~Longden,
Phys. Rev. D \textbf{94} (2016) no.2, 023506
doi:10.1103/PhysRevD.94.023506
[arXiv:1605.06350 [astro-ph.CO]].

\bibitem{Koh:2018qcy}
S.~Koh, B.~H.~Lee and G.~Tumurtushaa,
Phys. Rev. D \textbf{98} (2018) no.10, 103511
doi:10.1103/PhysRevD.98.103511
[arXiv:1807.04424 [astro-ph.CO]].

\bibitem{Watanabe:2006ku}
Y.~Watanabe and E.~Komatsu,
Phys. Rev. D \textbf{75} (2007), 061301
doi:10.1103/PhysRevD.75.061301
[arXiv:gr-qc/0612120 [gr-qc]].



\bibitem{Ellis:2021kad}
J.~Ellis, M.~A.~G.~Garcia, D.~V.~Nanopoulos, K.~A.~Olive and S.~Verner,
Phys. Rev. D \textbf{105} (2022) no.4, 043504
doi:10.1103/PhysRevD.105.043504
[arXiv:2112.04466 [hep-ph]].


\bibitem{Rinaldi:2015uvu}
M.~Rinaldi and L.~Vanzo,
Phys. Rev. D \textbf{94} (2016) no.2, 024009
doi:10.1103/PhysRevD.94.024009
[arXiv:1512.07186 [gr-qc]].

\bibitem{Amin:2014eta}
M.~A.~Amin, M.~P.~Hertzberg, D.~I.~Kaiser and J.~Karouby,
Int. J. Mod. Phys. D \textbf{24} (2014), 1530003
doi:10.1142/S0218271815300037
[arXiv:1410.3808 [hep-ph]].


\bibitem{Choi:2016eif}
S.~M.~Choi and H.~M.~Lee,
Eur. Phys. J. C \textbf{76} (2016) no.6, 303
doi:10.1140/epjc/s10052-016-4150-5
[arXiv:1601.05979 [hep-ph]].


\bibitem{Fei:2017fub}
Q.~Fei, Y.~Gong, J.~Lin and Z.~Yi,
JCAP \textbf{08} (2017), 018
doi:10.1088/1475-7516/2017/08/018
[arXiv:1705.02545 [gr-qc]].

\bibitem{Martin:2014nya}
J.~Martin, C.~Ringeval and V.~Vennin,
Phys. Rev. Lett. \textbf{114} (2015) no.8, 081303
doi:10.1103/PhysRevLett.114.081303
[arXiv:1410.7958 [astro-ph.CO]].


\bibitem{Gong:2015qha}
J.~O.~Gong, S.~Pi and G.~Leung,
JCAP \textbf{05} (2015), 027
doi:10.1088/1475-7516/2015/05/027
[arXiv:1501.03604 [hep-ph]].


\bibitem{Cai:2015soa}
R.~G.~Cai, Z.~K.~Guo and S.~J.~Wang,
Phys. Rev. D \textbf{92} (2015), 063506
doi:10.1103/PhysRevD.92.063506
[arXiv:1501.07743 [gr-qc]].

\bibitem{Ueno:2016dim}
Y.~Ueno and K.~Yamamoto,
Phys. Rev. D \textbf{93} (2016) no.8, 083524
doi:10.1103/PhysRevD.93.083524
[arXiv:1602.07427 [astro-ph.CO]].



\bibitem{Eshaghi:2016kne}
M.~Eshaghi, M.~Zarei, N.~Riazi and A.~Kiasatpour,
Phys. Rev. D \textbf{93} (2016) no.12, 123517
doi:10.1103/PhysRevD.93.123517
[arXiv:1602.07914 [astro-ph.CO]].



\bibitem{DeHaro:2017abf}
J.~De Haro and L.~Arest\'e Sal\'o,
Phys. Rev. D \textbf{95} (2017) no.12, 123501
doi:10.1103/PhysRevD.95.123501
[arXiv:1702.04212 [gr-qc]].

\bibitem{Artymowski:2017pua}
M.~Artymowski, O.~Czerwinska, Z.~Lalak and M.~Lewicki,
JCAP \textbf{04} (2018), 046
doi:10.1088/1475-7516/2018/04/046
[arXiv:1711.08473 [astro-ph.CO]].

\bibitem{Opferkuch:2019zbd}
T.~Opferkuch, P.~Schwaller and B.~A.~Stefanek,
JCAP \textbf{07} (2019), 016
doi:10.1088/1475-7516/2019/07/016
[arXiv:1905.06823 [gr-qc]].

\bibitem{Dimopoulos:2018wfg}
K.~Dimopoulos and T.~Markkanen,
JCAP \textbf{06} (2018), 021
doi:10.1088/1475-7516/2018/06/021
[arXiv:1803.07399 [gr-qc]].


\bibitem{Lankinen:2019ifa}
J.~Lankinen, O.~Kerppo and I.~Vilja,
Phys. Rev. D \textbf{101} (2020) no.6, 063529
doi:10.1103/PhysRevD.101.063529
[arXiv:1910.07520 [gr-qc]].


\bibitem{Hasegawa:2019jsa}
T.~Hasegawa, N.~Hiroshima, K.~Kohri, R.~S.~L.~Hansen, T.~Tram and S.~Hannestad,
JCAP \textbf{12} (2019), 012
doi:10.1088/1475-7516/2019/12/012
[arXiv:1908.10189 [hep-ph]].


\bibitem{Sadjadi:2012zp}
H.~M.~Sadjadi and P.~Goodarzi,
JCAP \textbf{02} (2013), 038
doi:10.1088/1475-7516/2013/02/038
[arXiv:1203.1580 [gr-qc]].

\bibitem{Braden:2010wd}
J.~Braden, L.~Kofman and N.~Barnaby,
JCAP \textbf{07} (2010), 016
doi:10.1088/1475-7516/2010/07/016
[arXiv:1005.2196 [hep-th]].

\bibitem{Nakayama:2008wy}
K.~Nakayama, S.~Saito, Y.~Suwa and J.~Yokoyama,
JCAP \textbf{06} (2008), 020
doi:10.1088/1475-7516/2008/06/020
[arXiv:0804.1827 [astro-ph]].

\bibitem{Harigaya:2013vwa}
K.~Harigaya and K.~Mukaida,
JHEP \textbf{05} (2014), 006
doi:10.1007/JHEP05(2014)006
[arXiv:1312.3097 [hep-ph]].


\bibitem{Tashiro:2003qp}
H.~Tashiro, T.~Chiba and M.~Sasaki,
Class. Quant. Grav. \textbf{21} (2004), 1761-1772
doi:10.1088/0264-9381/21/7/004
[arXiv:gr-qc/0307068 [gr-qc]].

\bibitem{Kofman:2005yz}
L.~Kofman and P.~Yi,
Phys. Rev. D \textbf{72} (2005), 106001
doi:10.1103/PhysRevD.72.106001
[arXiv:hep-th/0507257 [hep-th]].

\bibitem{Munoz:2014eqa}
J.~B.~Munoz and M.~Kamionkowski,
Phys. Rev. D \textbf{91} (2015) no.4, 043521
doi:10.1103/PhysRevD.91.043521
[arXiv:1412.0656 [astro-ph.CO]].

\bibitem{Dai:2014jja}
L.~Dai, M.~Kamionkowski and J.~Wang,
Phys. Rev. Lett. \textbf{113} (2014), 041302
doi:10.1103/PhysRevLett.113.041302
[arXiv:1404.6704 [astro-ph.CO]].

\bibitem{Allahverdi:2010xz}
R.~Allahverdi, R.~Brandenberger, F.~Y.~Cyr-Racine and A.~Mazumdar,
Ann. Rev. Nucl. Part. Sci. \textbf{60} (2010), 27-51
doi:10.1146/annurev.nucl.012809.104511
[arXiv:1001.2600 [hep-th]].

\bibitem{Odintsov:2022sdk}
S.~D.~Odintsov and V.~K.~Oikonomou,
[arXiv:2203.10599 [gr-qc]].


\bibitem{Hwang:2005hb}
J.~c.~Hwang and H.~Noh,
Phys. Rev. D \textbf{71} (2005), 063536
doi:10.1103/PhysRevD.71.063536
[arXiv:gr-qc/0412126 [gr-qc]].


\bibitem{Odintsov:2020sqy}
S.~D.~Odintsov, V.~K.~Oikonomou and F.~P.~Fronimos,
Nucl. Phys. B \textbf{958} (2020), 115135
doi:10.1016/j.nuclphysb.2020.115135
[arXiv:2003.13724 [gr-qc]].





\bibitem{Planck:2018jri}
Y.~Akrami \textit{et al.} [Planck],
Astron. Astrophys. \textbf{641} (2020), A10
doi:10.1051/0004-6361/201833887
[arXiv:1807.06211 [astro-ph.CO]].































\bibitem{LIGOScientific:2017vwq}
B.~P.~Abbott \textit{et al.} [LIGO Scientific and Virgo],
Phys. Rev. Lett. \textbf{119} (2017) no.16, 161101
doi:10.1103/PhysRevLett.119.161101
[arXiv:1710.05832 [gr-qc]].


\bibitem{LIGOScientific:2017ync}
B.~P.~Abbott \textit{et al.} [LIGO Scientific, Virgo, Fermi GBM, INTEGRAL, IceCube, AstroSat Cadmium Zinc Telluride Imager Team, IPN, Insight-Hxmt, ANTARES, Swift, AGILE Team, 1M2H Team, Dark Energy Camera GW-EM, DES, DLT40, GRAWITA, Fermi-LAT, ATCA, ASKAP, Las Cumbres Observatory Group, OzGrav, DWF (Deeper Wider Faster Program), AST3, CAASTRO, VINROUGE, MASTER, J-GEM, GROWTH, JAGWAR, CaltechNRAO, TTU-NRAO, NuSTAR, Pan-STARRS, MAXI Team, TZAC Consortium, KU, Nordic Optical Telescope, ePESSTO, GROND, Texas Tech University, SALT Group, TOROS, BOOTES, MWA, CALET, IKI-GW Follow-up, H.E.S.S., LOFAR, LWA, HAWC, Pierre Auger, ALMA, Euro VLBI Team, Pi of Sky, Chandra Team at McGill University, DFN, ATLAS Telescopes, High Time Resolution Universe Survey, RIMAS, RATIR and SKA South Africa/MeerKAT],
Astrophys. J. Lett. \textbf{848} (2017) no.2, L12
doi:10.3847/2041-8213/aa91c9
[arXiv:1710.05833 [astro-ph.HE]].


\bibitem{LIGOScientific:2017zic}
B.~P.~Abbott \textit{et al.} [LIGO Scientific, Virgo, Fermi-GBM and INTEGRAL],
Astrophys. J. Lett. \textbf{848} (2017) no.2, L13
doi:10.3847/2041-8213/aa920c
[arXiv:1710.05834 [astro-ph.HE]].


\bibitem{LIGOScientific:2018cki}
B.~P.~Abbott \textit{et al.} [LIGO Scientific and Virgo],
Phys. Rev. Lett. \textbf{121} (2018) no.16, 161101
doi:10.1103/PhysRevLett.121.161101
[arXiv:1805.11581 [gr-qc]].


\bibitem{LIGOScientific:2018hze}
B.~P.~Abbott \textit{et al.} [LIGO Scientific and Virgo],
Phys. Rev. X \textbf{9} (2019) no.1, 011001
doi:10.1103/PhysRevX.9.011001
[arXiv:1805.11579 [gr-qc]].



\bibitem{Ezquiaga:2017ekz}
J.~M.~Ezquiaga and M.~Zumalac\'arregui,
Phys. Rev. Lett. \textbf{119} (2017) no.25, 251304
doi:10.1103/PhysRevLett.119.251304
[arXiv:1710.05901 [astro-ph.CO]].


\bibitem{Sakstein:2017xjx}
J.~Sakstein and B.~Jain,
Phys. Rev. Lett. \textbf{119} (2017) no.25, 251303
doi:10.1103/PhysRevLett.119.251303
[arXiv:1710.05893 [astro-ph.CO]].


\bibitem{LIGOScientific:2018dkp}
B.~P.~Abbott \textit{et al.} [LIGO Scientific and Virgo],
Phys. Rev. Lett. \textbf{123} (2019) no.1, 011102
doi:10.1103/PhysRevLett.123.011102
[arXiv:1811.00364 [gr-qc]].




\bibitem{Oikonomou:2020sij}
V.~K.~Oikonomou and F.~P.~Fronimos,
Class. Quant. Grav. \textbf{38} (2021) no.3, 035013
doi:10.1088/1361-6382/abce47
[arXiv:2006.05512 [gr-qc]].

\bibitem{Odintsov:2020xji}
S.~D.~Odintsov, V.~K.~Oikonomou and F.~P.~Fronimos,
Annals Phys. \textbf{420} (2020), 168250
doi:10.1016/j.aop.2020.168250
[arXiv:2007.02309 [gr-qc]].











\bibitem{Venikoudis:2021irr}
S.~A.~Venikoudis and F.~P.~Fronimos,
Eur. Phys. J. Plus \textbf{136} (2021) no.3, 308
doi:10.1140/epjp/s13360-021-01298-y
[arXiv:2103.01875 [gr-qc]].


\bibitem{Venikoudis:2021oee}
S.~A.~Venikoudis and F.~P.~Fronimos,
Gen. Rel. Grav. \textbf{53} (2021) no.8, 75
doi:10.1007/s10714-021-02846-8
[arXiv:2107.09457 [gr-qc]].



\bibitem{Oikonomou:2021kql}
V.~K.~Oikonomou,
Class. Quant. Grav. \textbf{38} (2021) no.19, 195025
doi:10.1088/1361-6382/ac2168
[arXiv:2108.10460 [gr-qc]].

\bibitem{Odintsov:2020zkl}
S.~D.~Odintsov and V.~K.~Oikonomou,
Phys. Lett. B \textbf{805} (2020), 135437
doi:10.1016/j.physletb.2020.135437
[arXiv:2004.00479 [gr-qc]].


\bibitem{Kanti:2015pda}
P.~Kanti, R.~Gannouji and N.~Dadhich,
Phys. Rev. D \textbf{92} (2015) no.4, 041302
doi:10.1103/PhysRevD.92.041302
[arXiv:1503.01579 [hep-th]].


\bibitem{Fomin:2020hfh}
I.~Fomin,
Eur. Phys. J. C \textbf{80} (2020) no.12, 1145
doi:10.1140/epjc/s10052-020-08718-w
[arXiv:2004.08065 [gr-qc]].

\bibitem{Pozdeeva:2020apf}
E.~O.~Pozdeeva, M.~R.~Gangopadhyay, M.~Sami, A.~V.~Toporensky and S.~Y.~Vernov,
Phys. Rev. D \textbf{102} (2020) no.4, 043525
doi:10.1103/PhysRevD.102.043525
[arXiv:2006.08027 [gr-qc]].






\bibitem{Yi:2018gse}
Z.~Yi, Y.~Gong and M.~Sabir,
Phys. Rev. D \textbf{98} (2018) no.8, 083521
doi:10.1103/PhysRevD.98.083521
[arXiv:1804.09116 [gr-qc]].

\bibitem{Wu:2017joj}
Q.~Wu, T.~Zhu and A.~Wang,
Phys. Rev. D \textbf{97} (2018) no.10, 103502
doi:10.1103/PhysRevD.97.103502
[arXiv:1707.08020 [gr-qc]].


\bibitem{Rashidi:2020wwg}
N.~Rashidi and K.~Nozari,
Astrophys. J. \textbf{890}, 58
doi:10.3847/1538-4357/ab6a10
[arXiv:2001.07012 [astro-ph.CO]].


\bibitem{Odintsov:2020mkz}
S.~D.~Odintsov, V.~K.~Oikonomou, F.~P.~Fronimos and S.~A.~Venikoudis,
Phys. Dark Univ. \textbf{30} (2020), 100718
doi:10.1016/j.dark.2020.100718
[arXiv:2009.06113 [gr-qc]].

\bibitem{Oikonomou:2020tct}
V.~K.~Oikonomou and F.~P.~Fronimos,
Eur. Phys. J. Plus \textbf{135} (2020) no.11, 917
doi:10.1140/epjp/s13360-020-00926-3
[arXiv:2011.03828 [gr-qc]].


\bibitem{Oikonomou:2020oil}
V.~K.~Oikonomou and F.~P.~Fronimos,
EPL \textbf{131} (2020) no.3, 30001
doi:10.1209/0295-5075/131/30001
[arXiv:2007.11915 [gr-qc]].




\bibitem{Oikonomou:2019gjj}
V.~K.~Oikonomou and N.~T.~Chatzarakis,
Annals Phys. \textbf{411} (2019), 167999
doi:10.1016/j.aop.2019.167999
[arXiv:1908.09218 [gr-qc]].
 







\end{thebibliography}
\end{document}